\documentclass[prd,aps,tightenlines,superscriptaddress,nofootinbib,preprint]{revtex4}

\usepackage{graphicx}
\usepackage{dcolumn}
\usepackage{bm}
\usepackage{color}
\usepackage{amsmath}
\usepackage{subfigure}
\usepackage{amsfonts}
\usepackage{amssymb}
\usepackage{comment}
\usepackage{placeins}
\usepackage{orcidlink}
\begin{document}
\title{Geometric effects of torsion on black hole ringdown and shadows in Poincar\'e gauge gravity}

\author{Terkaa Victor Targema \orcidlink{0000-0001-8809-1741}}
\email{s2571502@ipc.fukushima-u.ac.jp; terkaa.targema@tsuniversity.edu.ng}
\affiliation{Division of Human Support System, Faculty of Symbiotic Systems Science, Fukushima University, Fukushima 960-1296, Japan}
\affiliation{Department of Physics, Taraba State University, Jalingo, Nigeria}

\author{Kazuharu Bamba \orcidlink{0000-0001-9720-8817}}
\email{bamba@sss.fukushima-u.ac.jp}
\affiliation{Division of Human Support System, Faculty of Symbiotic Systems Science, Fukushima University, Fukushima 960-1296, Japan}

\author{Riasat Ali \orcidlink{0000-0002-9371-1113}}
\email{riasatyasin@gmail.com}
\affiliation{Department of Mathematics, School of Science, University of Management and Technology,  Lahore, 54000, Pakistan}

\author{Usman Zafar \orcidlink{0000-0001-9610-1081}}
\email{s2471001@ipc.fukushima-u.ac.jp; zafarusman494@gmail.com}
\affiliation{Division of Human Support System, Faculty of Symbiotic Systems Science, Fukushima University, Fukushima 960-1296, Japan}

\begin{abstract}
Spacetime torsion provides a natural extension of general relativity and may lead to black hole solutions that differ significantly from their Einsteinian counterparts. We investigate a class of Reissner--Nordstr\"om-like black holes in Poincar\'e gauge gravity, where the effective charge is generated entirely by spacetime torsion instead of an electromagnetic field. Within the physically relevant torsion sector, the spacetime exhibits a single-horizon structure, free from the inner horizons and extremal states characteristic of charged black holes.
Using the sixth-order Wentzel-Kramers-Brillouin (WKB) approximation, Leaver's continued-fraction method, and the eikonal correspondence between quasinormal modes and unstable null geodesics, we study scalar perturbations and spin-2 test fields on the torsion-modified background. We find that increasing torsion decreases both the oscillation frequencies and damping rates, leading to longer-lived ringdown signals. 
We further compare the model predictions with Event Horizon Telescope observations of Sgr A* and M87*, along with representative LIGO–Virgo–KAGRA (LVK) ringdown scales, to derive constraints on the torsion parameter via a profile-$\chi^{2}$ analysis supplemented by Monte Carlo sampling. Although the resulting bounds remain consistent with the Schwarzschild limit within current observational uncertainties, our results show that spacetime torsion leaves correlated imprints on both black hole shadows and ringdown observables.
\end{abstract}

\maketitle











\section{Introduction}
\label{introduction}
The theory of general relativity (GR) offers a geometric characterization of gravitation, in which spacetime curvature arises from energy and momentum. Its predictions have been validated with extraordinary accuracy, from classical experiments in the solar system to the direct observation of gravitational waves from compact binary mergers \citep{iorio2020new,  schmidt2021machine, will2018new, abac2025gwtc, abbott2016observation, abbott2019gwtc, abbott2021gwtc}. Despite this success, several important questions remain open. These include the nature of dark matter and dark energy, the fate of spacetime singularities, and the problem of constructing a consistent quantum theory of gravity \citep{cirelli2026dark, bertone2005particle, clifton2012modified, r1, r2, ovgun2018tunneling, rovelli2004quantum, nojiri2011unified, ali2022quantum}. Such issues continue to motivate the study of extensions of GR, particularly in regimes involving strong gravitational fields or high energies. The study of black holes offers a robust framework for examining modified gravitational theories because features such as event horizons, photon spheres, and ringdown spectra encode valuable insights regarding the underlying gravitational structure. 

One possible direction is provided by Poincaré gauge gravity (PG), which is formulated within the Riemann-Cartan framework and allows for both curvature and torsion of spacetime. In this setting, gravitation is influenced not only by the energy-momentum tensor but also by the intrinsic spin of matter \citep{hehl2013poincare, p2}. The presence of torsion introduces additional geometric degrees of freedom and can modify the gravitational dynamics, especially when spin effects become relevant. In theories with dynamical torsion, such as PG theory, vacuum solutions need not coincide with those of GR, and even standard results such as Birkhoff's theorem can be violated in the presence of propagating torsion \citep{shapiro2002physical,obukhov2020generalized}. An interesting example is provided by the Reissner--Nordstr\"om (RN)-like black hole solution of Ref.~\citep{cembranos2017new}, in which the role of the electric charge is effectively replaced by a parameter generated entirely by spacetime torsion. As a result, the geometry closely resembles the RN spacetime while remaining a genuine vacuum solution of the underlying PG theory.

From an observational perspective, recent advances have opened new opportunities to probe gravity within the strong-field regime. In particular, the Event Horizon Telescope (EHT) has produced horizon-scale images of M87$^{*}$ and Sgr~A$^{*}$ (see Refs. \citep{Eht1,Eht2}), providing a powerful probe of black hole spacetimes. These observations have been widely employed to test GR and constrain alternative theories of gravity, although their interpretation remains dependent on astrophysical modeling of the accretion environment \citep{ca9,ca6,ca7,ca8}. This naturally motivates the search for complementary observables that are more directly linked to the underlying spacetime geometry.
The observational properties of torsion black holes have attracted considerable attention in recent years. For example, the effects of PG torsion on black hole shadows, weak gravitational lensing, and greybody factors were investigated in Ref.~\citep{pantig2023testing}, where torsion modifies a charged black hole geometry supported by a Maxwell-type electromagnetic field.  Related studies have also explored observational manifestations of torsion within teleparallel gravity in a variety of astrophysical contexts \citep{jusufi2022testing,bahamonde2020solar}.

Among the vacuum solutions admitted by PG gravity, those endowed with a purely geometric charge offer a particularly clean setting for isolating genuine torsion effects, free from the influence of electromagnetic fields \citep{blagojevic2022entropy}. Although this class of black hole solutions has been established theoretically, its phenomenological implications remain largely unexplored, particularly in light of contemporary observational probes. Motivated by this, we investigate whether a purely torsion-generated charge leaves observable signatures in black hole ringdown and shadow measurements. In this regard, recent observations by the LIGO-Virgo-KAGRA (LVK) collaboration have opened a complementary avenue for testing gravity through gravitational waves, with a rapidly growing catalogue of compact-binary merger events now available \citep{abbott2023gwtc,  acernese2015advanced,aso2013interferometer,collaboration2015advanced, abac2026directional, abac2026gwtc, abac2026searches}. Following a merger, the remnant black hole relaxes toward equilibrium by emitting gravitational radiation whose late-time ringdown signal is governed by a discrete spectrum of quasinormal modes (QNMs). The corresponding oscillation frequencies and damping times are determined solely by the background spacetime geometry, making the ringdown phase a sensitive probe of the strong-field gravitational field \citep{kokkotas1999quasi,berti2009quasinormal, abbott2025tests}. Consequently, departures from GR can manifest themselves as characteristic shifts in the QNM spectrum, and such signatures have been extensively investigated in a broad range of modified gravity theories \citep{konoplya2011quasinormal,Q2,Q3,Q4,Q5,Q6,Q7,Q8,fernando2012quasinormal,Q10,Q11,Q12,konoplya2026quasinormal, gong2024quasinormal, mishra2020quasinormal}. Motivated by these developments, we investigate the QNM spectrum of the torsion-generated RN-like black hole, focusing on scalar and spin-2 perturbations together with their interpretation through the eikonal correspondence between QNMs and unstable null geodesics. Complementing the ringdown analysis, we also compare the predicted shadow geometry with the EHT observations of M87* and Sgr A*, thereby deriving phenomenological constraints on the torsion parameter.

This paper is organized as follows; in Sec.~\ref{s2}, we study the black hole solution and the relevant geometrical aspects of PG gravity. In Sec.~\ref{s3}, we investigate the QNM frequencies using the Wentzel-Kramers-Brillouin (WKB) approximation, supplemented by the null-geodesic correspondence in the eikonal regime and validated through Leaver's method. We also discuss the potential observational distinctions between GR and PG gravity at LVK scales. In Sec.~\ref{s4}, we analyze the black hole shadow and compare the model with EHT observations. Finally, Sec.~\ref{s5} provides a concise overview of our results and suggests a possible path for further investigation.
\section{Torsion-induced Reissner--Nordstr\"om-like black hole}
\label{s2}
In this section, we briefly review the geometry and construction of the black hole solution, following Ref.~\citep{Blagojevic2019Entropy}. The PG theory describes gravity as a gauge theory of the localized Poincar\'e group of spacetime symmetries \citep{Obukhov2006PGT}. In this framework, the tetrad field, $\vartheta^{i}$ and the Lorentz connection, $\omega^{ij}$ act as gauge potentials \citep{BlagojevicHehl2013Gauge}. Their field strengths are given by the torsion,
\begin{equation}
\label{tor}
T^{i} = d\vartheta^{i} + \omega^{i}{}_{k}\,\vartheta^{k}\,,
\end{equation}
and the curvature
\begin{equation}
  \label{cuv}
  R^{ij} = d\omega^{ij} + \omega^{i}{}_{k}\,\omega^{kj}\,,
\end{equation}
so that the spacetime geometry is of the Riemann–Cartan type.
In a vacuum, the dynamics are encoded in a 4-form gravitational Lagrangian that depends on the tetrad, torsion, and curvature. Assuming parity invariance and terms at most quadratic in the field strengths, one may write the gravitational Lagrangian as 
\begin{eqnarray}
    L_{G} &= -^\star \left( a_{0} R + 2 \Lambda \right)
   + \sum_{n=1}^{3} a_{n}\, ^\star \left( {}^{(n)}T^{i} \wedge T_{i} \right) \nonumber \\
   &\quad + \frac{1}{2} \sum_{n=1}^{6} b_{n}\, ^\star \left( R^{ij} \wedge {}^{(n)}R_{ij} \right)\,,
\end{eqnarray}
with $(\Lambda, a_{0}, a_{n}, b_{n})$ is coupling constants, and ${}^{(n)}T^{i}$ and ${}^{(n)}R^{ij}$ denote the irreducible parts of torsion and curvature \citep{Obukhov2018PGT, Blagojevic2019Entropy}. Variation of $L_{G}$ with respect to $\vartheta^{i}$ and $\omega^{ij}$ yields the field equations in compact form,
\begin{eqnarray}
\nabla H_{i} + E_{i} &= 0\,, \\
\nabla H_{ij} + E_{ij} &= 0\,,
\end{eqnarray}
where the covariant momenta are
\[
H_{i} := \frac{\partial L_{G}}{\partial T^{i}}\,, 
\qquad
H_{ij} := \frac{\partial L_{G}}{\partial R^{ij}}\,,
\]
and the corresponding energy–momentum and spin currents are
\[
E_{i} := \frac{\partial L_{G}}{\partial \vartheta^{i}}\,, 
\qquad
E_{ij} := \frac{\partial L_{G}}{\partial \omega^{ij}}\,.
\]

We consider a static, spherically symmetric, and asymptotically flat spacetime with signature $(+,-,-,-)$, originally introduced in Ref.~\cite{cembranos2017new} and subsequently investigated in Ref.~\cite{blagojevic2022entropy}. The derivation of the corresponding torsion black hole solution, together with a detailed analysis of its geometric and thermodynamic properties, is presented therein. Additional background on the underlying framework of PG theories may also be found in Refs. \citep{grignani1992gravity, hayashi1980gravity, hehl2013poincare}. The resulting line element has the same functional structure as the RN geometry, with the effective charge term arising from torsion contributions. It is given by
\begin{equation}
ds^2 = f^2 dt^2 
  - \frac{dr^2}{f^2} 
  - r^2 \bigl(d\theta^2 + \sin^2\theta\, d\varphi^2 \bigr)\,,
\label{eq:metric}
\end{equation}
with lapse function
\begin{equation}\label{MF}
f^2(r)=1-\frac{2m}{r}+\frac{q}{r^2}\,,
\end{equation}
where \(m\) denotes the black hole mass and \(q\) characterizes the torsion-induced contribution that replaces the electromagnetic charge term of the standard RN solution. Consequently, Eq.~\eqref{MF} provides a simple and physically motivated framework for investigating the observational consequences of spacetime torsion. The horizon structure follows from the roots of $f^2(r)=0$, which are given by
\begin{equation}
r^2 - 2mr + q = 0\,,
\end{equation}
with solutions
\begin{equation}
r_{\pm} = m \pm \sqrt{m^2 - q}\,.
\label{eq:roots}
\end{equation}

This horizon structure exhibits several distinct features compared with the classical black holes in GR. For $q>0$ and $q\leq m^2$, the geometry admits two real roots corresponding to inner and outer horizons. If $q>m^2$, the horizons disappear and the central singularity becomes naked. The regime $q<0$ is unique to this torsion-induced RN-like solution. As the magnitude of the negative torsion parameter increases, the horizon structure is progressively modified, and beyond a critical threshold the Cauchy horizon disappears. Consequently, the spacetime no longer admits the horizon degeneracy associated with the extremal limit of charged black holes in GR. In the following, to clarify the origin of the parameter $q$ and its impact on the horizon structure, we briefly review the underlying torsion configuration.

A convenient orthonormal co-frame is chosen as
\begin{eqnarray}
    \vartheta^0 = f\, dt\,, 
\qquad 
\vartheta^1 = \frac{dr}{f}, \nonumber\\
\vartheta^2 = r\, d\theta\,, 
\qquad 
\vartheta^3 = r\sin\theta\, d\varphi\,,
\end{eqnarray}
which yields the horizon area
\begin{equation}
A_H = \int_{S^2} \vartheta^2 \wedge \vartheta^3 
    = 4\pi r_+^2 \,.
\end{equation}
Within the PG framework, the torsion 2-forms are assumed to take the static, spherically symmetric ansatz
\begin{eqnarray}
T^0 &=& T^1 = A\, \vartheta^0 \wedge \vartheta^1\,,  \\[4pt]
T^2 &=& - G\, \vartheta^- \wedge \vartheta^2 
       - H\, \vartheta^- \wedge \vartheta^3\,,  \\[4pt]
T^3 &=& - G\, \vartheta^- \wedge \vartheta^3 
       + H\, \vartheta^- \wedge \vartheta^2\,, 
\end{eqnarray}
with $\vartheta^- := \vartheta^0 - \vartheta^1$ and  \(A\), \(G\), and \(H\) are radial functions (see Ref. \citep{cembranos2017new}) given by
\begin{equation}
A = -\partial_r f\,,
\qquad
G = \frac{f}{2r}\,,
\qquad
H = \frac{p}{rf}\,,
\end{equation}
where $p$ is a new torsion parameter, complementing $q$ in the metric sector. This choice excites all irreducible pieces of torsion.

The Cartan structure Eq.~\eqref{tor} determines the Lorentz connection, with independent nonvanishing components
\begin{eqnarray}
\omega^{01} &=& - (\partial_r f)\, \vartheta^1\,, \\[4pt]
\omega^{0c} &=& - \omega^{1c} = - \frac{f}{2r}\, \vartheta^c\,, \qquad (c=2,3)\,, \\[4pt]
\omega^{23} &=& \frac{\cos\theta}{r\sin\theta}\,\vartheta^3 
               + \frac{p}{fr}\, \vartheta^- \,.
\end{eqnarray}
By using  Eq.~\eqref{cuv}, the curvature two-forms follow as:
\begin{eqnarray}
R_{A2} &=& -\frac{p}{2r^3}\, \vartheta^- \wedge \vartheta^3\,,  \\[4pt]
R_{A3} &=& \phantom{-}\frac{p}{2r^3}\, \vartheta^- \wedge \vartheta^2\,,  \\[4pt]
R_{23} &=& \frac{1}{r^2}\Bigl(p\, \vartheta^0 \wedge \vartheta^1 
            - \vartheta^2 \wedge \vartheta^3\Bigr)\,,
\end{eqnarray}
with $A=(0,1)$. Thus the pair $(\vartheta^i,\omega^{ij})$, parametrized by $(q,p)$, 
fully characterizes the torsion-modified RN-like geometry. 

From these quantities, one constructs the associated topological densities. The Euler density vanishes, whereas the Nieh--Yan and Pontryagin densities receive nontrivial torsional contributions (see Refs.~\cite{hehl1995metric, nieh1982identity}), which take the form
\begin{eqnarray}
I_E &=& \varepsilon_{ijmn}\, R^{mn} \wedge R^{ij} = 0\,,  \\[6pt]
I_{NY} &=& T^i \wedge T_i - R^{ij} \wedge \vartheta_i \wedge \vartheta_j 
        = \frac{2p}{r^2}\,\hat\epsilon\,,  \\[6pt]
I_P &=& R^{ij} \wedge R_{ij} 
        = - \frac{4p}{r^4}\,\hat\epsilon\,,
\end{eqnarray}
where $\hat\epsilon := \vartheta^0 \wedge \vartheta^1 
\wedge \vartheta^2 \wedge \vartheta^3$ is the invariant volume 4-form. The singularity at $r=0$ remains, but $p$ gains a clear topological role. 

The ansatz $(\vartheta^i,\omega^{ij})$ solves the PG field equations provided the Lagrangian couplings satisfy
\begin{eqnarray}
    &\Lambda = 0\,, \qquad a_1 - a_0 = 0\,, \qquad a_2 + 2a_0 = 0\,, \nonumber \\ 
    &a_3 + \frac{1}{2}a_0 = 0\,, \quad b_1 = b_4 = b_6 = 0\,, \nonumber \\
    &b_3 = -b_2\,, \qquad b_5 = -\frac{1}{3}b_2\,, 
\end{eqnarray}
so that the parameters $q$ and $p$ are dynamically related by
\begin{equation}\label{kp}
q = k p^2\,, 
\qquad k := -\frac{b_2}{3a_0}\,.
\end{equation}
It can be observed from Eq.~\eqref{kp} that the effective charge parameter does not constitute an independent degree of freedom; rather, it arises dynamically from the torsional structure of the spacetime.  This reduction simplifies the PG Lagrangian to
\begin{eqnarray}
L_G &=& - a_0 \, \star R 
    + a_0 \, T^i \star \Big( {}^{(1)}T_i - 2\, {}^{(2)}T_i - \frac{1}{2}{}^{(3)}T_i \Big) \nonumber \\
    && + \frac{1}{2} b_2 R^{ij} \star \Big( {}^{(2)}R_{ij} - {}^{(3)}R_{ij} - \frac{1}{3} {}^{(5)}R_{ij} \Big)\,,
\end{eqnarray}
with $a_0>0$ ensuring a smooth GR limit.  

\vspace{0.3cm}

The relation \(q=kp^{2}\) illustrates how torsion controls the horizon structure of the solution. This relation provides a direct association between torsion and the observable characteristics of the black hole spacetime, such as the horizon configuration, photon sphere, shadow size, and QNM spectrum, studied in this work. As shown in Ref.~\citep{ref17}, positivity of the spin-1 sector of the particle spectrum requires \(b_{2}>0\), which implies \(k<0\). Consequently, the physically admissible sector corresponds to \(q<0\). In this regime, the outer horizon \(r=r_{+}\) remains positive, whereas the inner root \(r_{-}\) is shifted to negative radius and therefore does not represent a physical Cauchy horizon. The spacetime thus possesses a single physical event horizon, while the torsion parameter \(p\) remains unrestricted. By contrast, the RN-like branch with \(q>0\), including configurations that admit naked singularities, is dynamically disfavored by the positivity condition. For clarity, the cases with \(q\geq0\) considered later in this work are not derived from the physically admissible PG branch selected by the positivity condition. Instead, they correspond to the general-relativistic Schwarzschild \((q=0)\) and RN-like \((q>0)\) geometries and are included solely as benchmarks for comparison with the PG framework. Accordingly, any results presented for \(q\geq0\) should not be interpreted as predictions of the physically admissible PG theory.

In the following, we incorporate the effects of torsion through the physically motivated choice \(q<0\) (equivalently \(k<0\)). A notable consequence is that the spacetime admits no extremal configuration. Moreover, unlike the RN solution, the torsion parameter is not constrained by a charge-to-mass bound imposed by cosmic censorship. For convenience, we denote \(f^{2}(r)\equiv F(r)\) throughout the remainder of this work.
\section{Quasinormal modes}
\label{s3}
In this section, we investigate the influence of spacetime torsion on the quasinormal-mode spectrum of the black hole. Our primary analysis is based on the sixth-order WKB approximation with Pad\'e resummation. To gain further physical insight into the underlying mechanism, we also employ the eikonal geodesic correspondence, which relates the quasinormal spectrum to the properties of unstable null geodesics. Finally, Leaver's continued-fraction method is used as an independent numerical benchmark to assess the accuracy and reliability of the WKB results.
\subsection{The WKB method}
To probe the dynamical response of the black hole spacetime, we study its QNM spectrum using the WKB method, originally introduced in Ref.~\citep{schutz1985black}. Throughout this analysis, we consider massless test-field perturbations propagating on the static, spherically symmetric background of Eq.~\eqref{eq:metric}. The evolution of the perturbing field is governed by the massless Klein-Gordon equation
\begin{equation}
\Box \Phi = 0.
\end{equation}
After separation of variables in the usual form
\begin{equation}
\Phi(t,r,\theta,\phi)=e^{-i\omega t}\,\frac{\psi(r)}{r}\,Y_{l m}(\theta,\phi)\,,
\end{equation}
the radial equation can be written in a simplified form by introducing the tortoise coordinate \(r_*\), defined through 
\(
\frac{dr_*}{dr}=\frac{1}{F(r)}\,,
\)
such that the corresponding wave equation can be cast into the Schr\"odinger-like form
\begin{equation}
\label{ess}
\frac{d^2\psi}{dr_*^2}+\left[\omega^2-\rm V_{\rm eff}(r)\right]\psi=0\,,
\end{equation}
with the effective potential
\begin{equation}
{\rm V}_{\rm eff}(r)=F(r)\left[\frac{(1-s^2)F'(r)}{r}
+\frac{l(l+1)}{r^2}\right],
\end{equation}
where $l=s,s+1,s+2,\cdots$ is the spherical harmonics index,  $s=0,1,2$ correspond to scalar, electromagnetic, and spin-2 perturbations, respectively. In the Schwarzschild (GR) limit, the \(s=2\) case reproduces the well-known Regge-Wheeler potential governing axial gravitational perturbations \citep{toshmatov2015quasinormal}. In the present torsion-based framework, however, additional propagating degrees of freedom associated with torsion may be present, and the physical gravitational perturbations can only be determined through a linearization of the full PG field equations. Consequently, the $s=2$ sector considered here corresponds to an effective massless spin-2 field propagating on the torsion-modified background, although it may not necessarily coincide with the axial gravitational perturbation of the underlying theory. Nevertheless, the resulting effective potential captures how the underlying geometry influences the propagation and temporary trapping of spin-2 waves near the photon sphere, thereby providing a useful phenomenological probe of the corresponding ringdown characteristics.  Since $F(r)$ is directly affected by the torsion parameter $(q)$, variation in the torsional sector modifies the effective potential, which leads to an impact on the QNM spectrum. Furthermore, QNM are defined by the physically motivated boundary conditions of purely ingoing waves at the event horizon and purely outgoing waves at spatial infinity \citep{hosseinifar2025quasinormal, churilova2020quasinormal}
\begin{equation}
\label{bcs}
\psi(r_*) \sim
\begin{cases}
e^{-i\omega r_*}, & r_* \to -\infty,\\[1ex]
e^{+i\omega r_*}, & r_* \to +\infty.
\end{cases}
\end{equation}
These conditions lead to a discrete set of complex frequencies \(\omega\), whose real part gives the oscillation frequency and whose imaginary part determines the damping rate.

To compute the spectrum, we use the WKB approximation around the maximum of the effective potential. Let \(r_0\) denote the position of the peak, defined by \(\rm V'_{\rm eff}(r_0)=0\), and let \(V_0=\rm V_{\rm eff}(r_0)\). Expanding the potential about \(r_0\) and applying the higher-order WKB construction yields the quantization condition
\begin{equation}
\frac{i\left(\omega^2-V_0\right)}{\sqrt{-2V_0''}}
-\sum_{k=2}^{N}\Lambda_k
= n+\frac{1}{2}\,,
\end{equation}
where \(n\) is the overtone number, \(V_0''\) denotes the second derivative with respect to \(r_*\) evaluated at the peak, and \(\Lambda_k\) are the higher-order WKB correction terms, which are detailed in Refs. \citep{fernando2012quasinormal, iyer1987black, konoplya2003quasinormal}. Accordingly, the QNMs has the form given by
\begin{equation}
    \omega={\rm Re}[\omega]-i{\rm Im}[\omega] \,.
    \end{equation}
    
This work employs the WKB expansion extended to sixth order. Since the raw WKB series may converge slowly, especially away from the eikonal regime, we improve the approximation by constructing a Pad\'e approximant of the truncated series. The resulting rational approximation stabilizes the WKB expression and typically yields a more accurate estimate of the complex quasinormal frequency. The final quasinormal frequencies are obtained by solving the Pad\'e-improved WKB condition (see Ref. \citep{konoplya2019higher}) for \(\omega\). Throughout this work, we restrict attention to scalar perturbations and spin-2 test fields only.
\begin{figure}[t]
\centering
\includegraphics[width=0.75\textwidth]{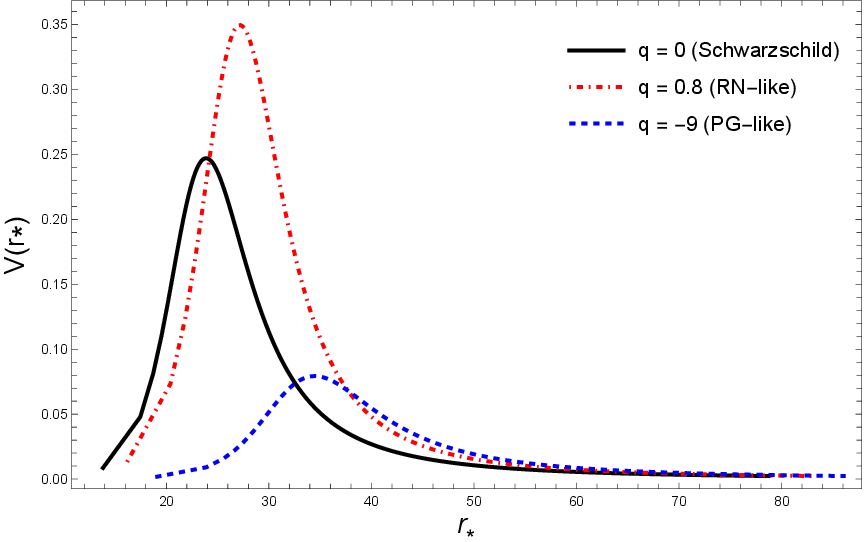}
\caption{
Effective potential profiles for scalar perturbations with multipole number $l=2$ for different values of the parameter $q$. The black solid curve corresponds to the Schwarzschild case ($q=0$), the red dot-dashed curve represents the RN-like regime ($q>0$), and the blue dashed curve denotes the torsion-dominated regime ($q<0$). Positive values of $q$ increase the height of the effective potential barrier, whereas negative values significantly suppress it, leading to distinct quasinormal-mode characteristics.
}
\label{effective_potential}
\end{figure}

In addition to the WKB results discussed above, it is instructive to consider the eikonal limit, which provides a complementary and physically transparent approximation to the quasinormal spectrum. In this regime, corresponding to large angular momentum \(l \gg n\), the effective potential simplifies and becomes dominated by the centrifugal term. Equation~(\ref{ess}) can be written as
\begin{equation}
\frac{d^2\psi}{dr_*^2} + \mathcal{Q}\,\psi=0\,, \qquad 
\mathcal{Q}=\omega^2 - {\rm V}_{\rm eff}(r)\,,
\end{equation}
with
\(
{\rm V}_{\rm eff}(r)\approx  F(r)\frac{l^{2}}{r^2}\,.
\)
In this limit, the maximum of the effective potential is positioned at the radius of unstable circular null geodesics, \(r_0\), which satisfies
\begin{equation}
\label{null}
\frac{d}{dr}\left(\frac{F(r)}{r^2}\right)\Big|_{r=r_0}=0\,.
\end{equation}

The maximum of the effective potential coincides with the location of the unstable null geodesics, and the WKB quantization condition as derived in Refs. \cite{cardoso2009geodesic, ali2025evaluation} becomes
\begin{equation}
\label{eqn:condw}
\frac{\mathcal{Q}(r_0)}{\sqrt{-2\frac{d^2\mathcal{Q}(r_0)}{dr_*^2}}}=i\left(n+\frac{1}{2}\right)\,,
\end{equation}
where $n$ is the overtone number. In the large-$l$ regime, the corresponding eikonal relation takes the form
\begin{equation*}
\omega=\left(l+\frac{1}{2}\right)\,\Omega_0-i\left(n+\frac{1}{2}\right)\lambda\,,
\end{equation*}
where \(\Omega_0\) is the angular frequency of the unstable null orbit and \(\lambda\) is the associated Lyapunov exponent, which quantifies the instability timescale of the orbit \citep{cardoso2009geodesic}.

For a general static, spherically symmetric spacetime defined in Eq~\eqref{eq:metric}, the angular frequency is given by
\begin{equation*}
\Omega_0=\frac{\dot{\phi}}{\dot{t}}=\sqrt{\frac{F(r_0)}{r_0^2}}\,,
\end{equation*}
while the Lyapunov exponent follows from the second derivative of the effective radial potential governing null geodesics. Restricting to null motion,  one finds the radial equation in terms of the conserved quantities as
\begin{equation}
\dot{r}^2 + {\rm V}_{\rm eff}=0\,, \qquad 
{\rm V}_{{\rm eff}}
=
F(r)\left[
\frac{{\rm L}^2}{r^2}
-
\frac{{\rm E}^2}{F(r)}
\right].
\end{equation}
where $\rm L$ and $\rm E$ are the angular momentum and particle energy, respectively. The corresponding Lyapunov exponent is given by 
\begin{equation}
\lambda=\sqrt{\frac{-\rm V_{\rm eff}''(r_0)}{2\,\dot{t}^2}}\;\,,
\end{equation}
which, for the metric under consideration, can be expressed explicitly as
\begin{equation}
\lambda=\sqrt{\frac{F(r_0)}{2r_0^2}\left[2F(r_0)-r_0^2 F''(r_0)\right]}\, .
\end{equation}

\begin{table*}[t]
\centering
\caption{QNMs for $l=2$, $n=0$. Here, we compare the eikonal (null-geodesic) approximation and the 6th-order WKB method for scalar perturbations.  A systematic decrease in both the oscillation frequency and damping rate is observed as $q$ becomes more negative, indicating the influence of torsion on the ringdown spectrum.}
\label{tab:qnm_l2}

\renewcommand{\arraystretch}{1.15}
\setlength{\tabcolsep}{10pt}

\begin{tabular}{c c c c c c c c}
\hline\hline
$q$
& \multicolumn{2}{c}{Geodesics}
& \multicolumn{2}{c}{WKB method}
& \multicolumn{2}{c}{Leaver Method}
& WKB error \\

& ${\rm Re}[\omega]$
& $-{\rm Im}[\omega]$
& ${\rm Re}[\omega]$
& $-{\rm Im}[\omega]$
& ${\rm Re}[\omega]$
& $-{\rm Im}[\omega]$
& $\Delta E\%$
\\
\hline

0.0   & 0.481125  & 0.096225  & 0.483589 & 0.0967142 & 0.483644 & 0.096759 & 0.014382 \\
0.2   & 0.498444 & 0.0972209 & 0.500724 & 0.0977705 & 0.501136  & 0.097733 & 0.081027 \\
0.4   & 0.518957  & 0.0979496 & 0.521792 & 0.0983893 & 0.521844 & 0.098429 & 0.012310 \\
0.6   & 0.544131 & 0.0980485 & 0.547164 & 0.0984383 & 0.547214 & 0.098475 & 0.011155 \\
0.8   & 0.576875  & 0.0964726 & 0.579982 & 0.0967722 & 0.580028 & 0.096803 & 0.009414 \\
\hline

-1.5  & 0.400080 & 0.0879708 & 0.40175  & 0.0884994 & 0.401809 & 0.088557 & 0.020040 \\
-3.0  & 0.355079  & 0.0814009 & 0.356367 & 0.0819239 & 0.356425 & 0.081988 & 0.023636 \\
-4.5  & 0.324506 & 0.0762960 & 0.325565 & 0.0768061 & 0.325622 & 0.076873 & 0.026269 \\
-6.0  & 0.301693  & 0.0721865 & 0.302596 & 0.0726825 & 0.302652 & 0.072752 & 0.028673 \\
-7.5  & 0.283695 & 0.0687769 & 0.284485 & 0.0692591 & 0.284539 & 0.069329 & 0.030161 \\
-9.0  & 0.268957 & 0.0658808 & 0.26966  & 0.0663501 & 0.269713 & 0.066421 & 0.031868 \\
-30.0 & 0.177956  & 0.0461385 & 0.178244 & 0.0464971 & 0.178284 & 0.046562 & 0.041374 \\
\hline\hline
\end{tabular}
\end{table*}
The geodesic correspondence provides a useful framework for identifying the common effects of torsion on quasinormal modes and black hole shadows. Since both the eikonal quasinormal spectrum and shadow observables are computed by the characteristics of unstable null geodesics, their combined analysis allows one to uncover unified signatures of torsion in black hole phenomenology. To further validate the quasinormal-mode results, we employ the Leaver (Frobenius) method in the next section as an independent and highly accurate benchmark for the sixth-order WKB calculations.
\subsection{Leaver method}

Among the available techniques for computing QNM frequencies, the continued-fraction method developed by Leaver~\cite{leaver1985analytic} is widely regarded as one of the most accurate and reliable. Owing to its rapid convergence and high numerical precision, it has become the standard criterion for confirming approximate approaches such as the WKB approximation and time-domain evolution. 

For neutral scalar perturbations of a static, spherically symmetric black hole, the radial solution satisfying the QNM boundary conditions \eqref{bcs} can be written in the Frobenius form
\begin{equation}
\label{eq:LeaverAnsatz}
R(r)=
r\,(r-r_-)^{-1+i\omega(r_++r_-)}
\left(
\frac{r-r_+}{r-r_-}
\right)^{-\frac{i\omega r_+^{\,2}}{r_+-r_-}}
e^{i\omega r}
\sum_{n=0}^{\infty}
a_n
\left(
\frac{r-r_+}{r-r_-}
\right)^n \, ,
\end{equation}
where $r_+$ and $r_-$ denote the outer and inner horizons, respectively, and the expansion automatically incorporates the required ingoing behavior at the event horizon and outgoing behavior at spatial infinity.

For the black hole considered in this work, the effects of torsion enter only through the parameter
$q=k p^{2}$ appearing in the metric function. Consequently, the mathematical structure of the radial equation closely parallels that of the RN spacetime (see Refs. \citep{berti2003asymptotic, leaver1990quasinormal}), allowing the continued-fraction formalism to be implemented with only the corresponding replacement of the horizon radii. Substituting Eq.~(\ref{eq:LeaverAnsatz}) into the radial wave equation yields the following three-term recurrence relations \citep{bonanno2025regular, konoplya2011quasinormal}:
\begin{align}
a_{1}\alpha_0  + a_{0}\beta_0  &= 0, \\
a_{n+1}\alpha_n  + a_n\beta_n  + a_{n-1}\gamma_n  &= 0,
\qquad n \ge 1.
\end{align}
with the coefficients given by
\begin{align}
\alpha_n &=
-(n+1)
\left[
(n+1)r_-
+r_+\left(-n+2i r_+\omega-1\right)
\right], \\[1ex]
\beta_n &=
-r_+
\left[
l(l+1)
+2n^2+2n+1
-4i(2n+1)r_+\omega
-8r_+^2\omega^2
\right]
\nonumber\\
&\quad
+r_-
\left[
l(l+1)
+2n(n+1)+1
\right]
-2i(2n+1)r_-r_+\omega, \\[1ex]
\gamma_n &=
-\left[
n-2i\omega(r_-+r_+)
\right]
\left[
n(r_--r_+)+2i r_+^{\,2}\omega
\right].
\end{align}

The quasinormal frequencies are obtained by imposing the corresponding infinite continued-fraction condition associated with the above recurrence relations. 
\subsection{Quasi-normal modes: results and discussions}
\begin{figure}[t]
    \centering
    \includegraphics[width=0.75\textwidth]{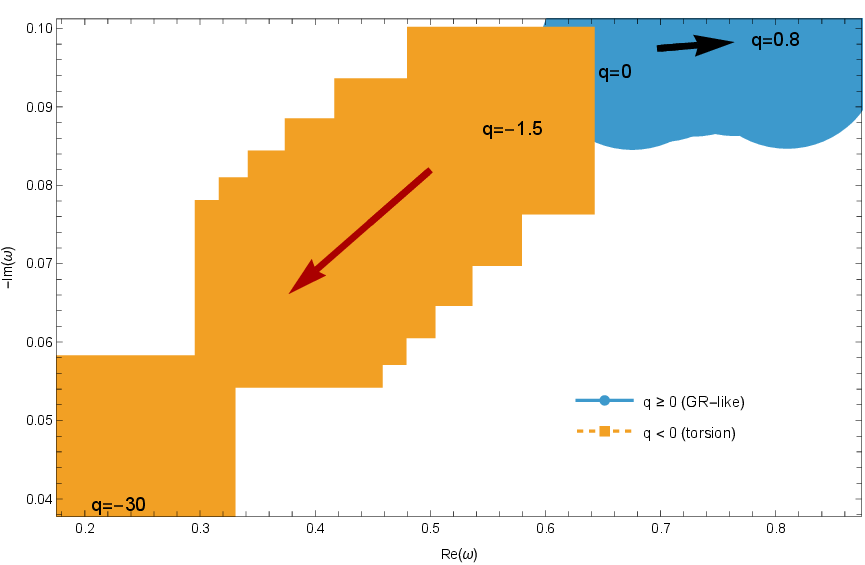}
    \caption{
Parametric evolution of the fundamental QNM (\(n=0\)) in the complex-frequency plane for scalar perturbations with \(l=3\). The blue branch corresponds to the general relativistic (GR-like) regime (\(q\geq0\)), while the orange branch represents the torsion-dominated regime (\(q<0\)). The arrows indicate the direction of parameter variation along each trajectory. Along the GR-like branch, increasing positive \(q\) produces only minor changes in the quasinormal frequencies. In contrast, increasing negative \(q\), corresponding to stronger torsion effects, drives the spectrum toward lower values of both \(\mathrm{Re}[\omega]\) and \(-\mathrm{Im}[\omega]\). This simultaneous reduction in the oscillation frequency and damping rate implies longer-lived ringdown modes. The clear separation between the two branches highlights the distinct imprint of torsion on the black hole QNM spectrum.
}  \label{fig:qnm_trajectory_torsion}
\end{figure}

The QNM spectra reported in Tables~\ref{tab:qnm_l2} to~\ref{tab:high_modes_torsion} demonstrate a clear and systematic dependence on the parameter $q$. In the $q>0$ regime, which corresponds to the RN limit, both the oscillation frequency and the damping rate are larger, reflecting a steeper effective potential barrier as shown in Fig. \ref{effective_potential}\footnote{This deformation of the potential profile modifies the curvature near its maximum, which plays a central role in semi-analytic approaches such as the WKB method. In particular, a flatter peak reduces the accuracy of higher-order WKB expansions, providing a natural explanation for the mild increase in WKB error observed in the torsion-dominated regime (see Table \ref{tab:qnm_l3_updated}). Nevertheless, the method remains robust, capturing the qualitative and quantitative trends of the quasinormal spectrum across both regimes.}.
As $q$ decreases and becomes negative, torsion effects in the PG geometry become increasingly important, and both ${\rm Re}[\omega]$ and $|{\rm Im}[\omega]|$ decrease monotonically. The resulting modes therefore oscillate more slowly and decay more gradually, indicating longer-lived perturbations (see Fig. \ref{fig:qnm_trajectory_torsion}). Physically, this behavior originates from the torsion-derived modification of the effective potential, as more negative values of $q$ displace the potential peak and smooth its profile. The flattening of the potential barrier weakens the restoring force near the photon sphere, resulting in ringdown modes at lower frequencies and longer damping times than in GR. 

The superior performance of the WKB method is expected when $l\gg n$. In the GR-dominated regime ($q>0$), the effective potential retains a sharp single-barrier structure across the physically allowed range, so the WKB approximation remains highly accurate and the associated errors are negligible, as seen in Table~\ref{tab:high_modes_torsion}. However, in the torsion-dominated regime the potential barrier becomes progressively flatter (Fig.~\ref{effective_potential}), reducing the accuracy of the WKB approximation and resulting in larger Pad\'e convergence errors\footnote{The Pad\'e convergence error is estimated from the difference between successive Pad\'e approximants.} even when $l\gg n$. Since $q$ is not tightly bounded from below in this geometry, sufficiently negative produces lower oscillations and weaker damping, and this is a major discrepancy between the GR and PG regimes. 
\begin{table*}[t]
\caption{
Fundamental (\(n=0\)) quasinormal frequencies for the spin-\(2\) (\(s=2\)) field, computed using the sixth-order WKB method  for different values of the torsion parameter \(q\). 
}
\label{tab:s2}

\begin{ruledtabular}
\begin{tabular}{c c c c c c c}

& \multicolumn{3}{c}{$l=2$}
& \multicolumn{3}{c}{$l=3$} \\

\hline
$q$
& Re$[\omega]$
& -Im$[\omega]$
& Pad\'e Error
& Re$[\omega]$
& -Im$[\omega]$
& Pad\'e Error \\

\hline

$0$
& 0.373622 & 0.0889866 & 0.135
& 0.599421 & 0.0926911 & 0.026 \\

$0.2$
& 0.389747 & 0.0899630 & 0.130
& 0.622922 & 0.0937715 & 0.025 \\

$0.4$
& 0.409269 & 0.0907082 & 0.121
& 0.6510398 & 0.0946065 & 0.023 \\

$0.6$
& 0.433935 & 0.0908617 & 0.108
& 0.686007 & 0.0948259 & 0.021 \\

$0.8$
& 0.467404 & 0.0892666 & 0.092
& 0.732358 & 0.0933072 & 0.022 \\

$-1.0$
& 0.319752 & 0.0836673 & 0.257
& 0.519222 & 0.0868477 & 0.034 \\

$-1.5$
& 0.301769 & 0.0812507 & 0.310
& 0.491902 & 0.0842109 & 0.037 \\

$-2.0$
& 0.287049 & 0.0790486 & 0.359
& 0.469345 & 0.0818187 & 0.040 \\

$-2.5$
& 0.274656 & 0.0770417 & 0.404
& 0.450218 & 0.0796471 & 0.043 \\

$-2.9$
& 0.266013 & 0.0755607 & 0.436
& 0.436809 & 0.0780499 & 0.045 \\

\end{tabular}
\end{ruledtabular}

\end{table*}

In Table~\ref{tab:s2}, we present the QNM frequencies of a massless spin-2 test field propagating on the black hole background. Although this field does not represent the full gravitational perturbation spectrum of the underlying PG theory, it serves as a useful probe of how the torsion-modified geometry influences the propagation of spin-2 disturbances. The results reveal a clear and systematic trend: increasing torsion suppresses both the oscillation frequency and the damping rate, producing longer-lived modes.

\begin{figure*}[t]
\centering

\begin{minipage}[t]{0.48\textwidth}
\centering
\includegraphics[width=\linewidth]{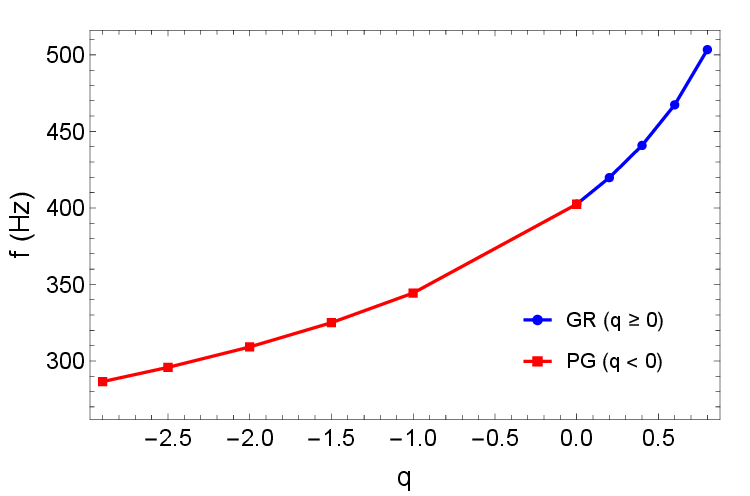}
\\[-2mm]
(a) 
\end{minipage}
\hfill
\begin{minipage}[t]{0.48\textwidth}
\centering
\includegraphics[width=\linewidth]{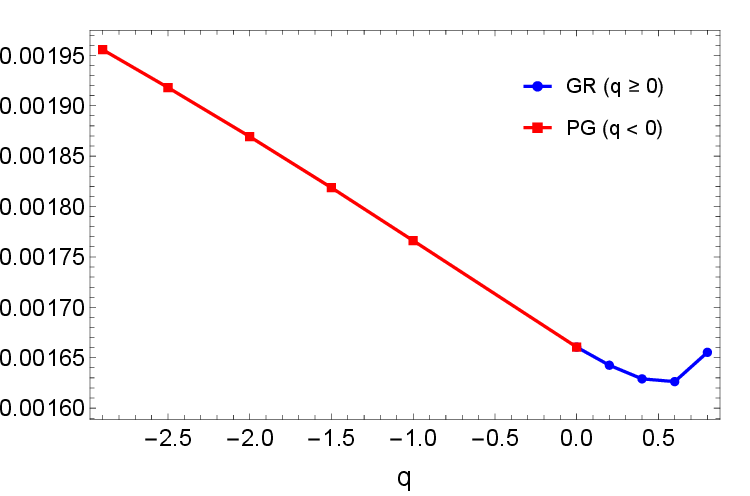}
\\[-2mm]
(b)
\end{minipage}

\caption{Effects of the torsion parameter $q$ on the ringdown frequencies and damping times of a massless spin-2 test field for a black hole with $M=30\,M_{\odot}$, representative of the mass scale probed by the LVK collaboration. The torsion-dominated regime (red curve) shifts the spectrum toward lower frequencies and longer damping times, whereas the GR-like regime (blue curve) exhibits the opposite trend.}\label{LVK-lisa}
\end{figure*}

To assess the phenomenological significance of this behavior, Fig.~\ref{LVK-lisa} displays the corresponding ringdown frequencies and damping times for a representative stellar-mass black hole of mass $M=30\,M_{\odot}$, chosen to lie within the mass range commonly probed by the LVK collaboration \citep{abbott2016tests,giesler2019black}.  A particularly interesting feature of these results is their geometric origin. The suppression of the quasinormal spectrum can be traced to the same torsion-induced deformation of the spacetime that alters the properties of unstable null geodesics. As we shall see in the following section, this mechanism also affects the photon-sphere structure and, consequently, the black hole shadow. This connection provides a natural bridge between ringdown and shadow observables as complementary probes of the torsional sector of PG theory. The dimensionless quasinormal frequencies are converted to physical observables through the standard relations 
\begin{equation}
f=\frac{\mathrm{Re}[\omega]c^{3}}{2\pi GM},
\qquad
\tau=\frac{GM}{c^{3}\left|\mathrm{Im}[\omega]\right|},
\end{equation}
where $f$ denotes the ringdown frequency and $\tau$ the damping time \citep{kokkotas1999quasi,berti2009quasinormal}.  Although based on a spin-2 test field, these results suggest that torsion may leave correlated signatures across multiple strong-gravity observables, motivating the complementary constraints on the torsion parameter derived in the next section from EHT observations of M87$^{*}$ and Sgr~A$^{*}$.
\section{EHT constraints on torsion}
\label{s4}
Black hole shadow observations of the supermassive black holes Sgr~A$^{*}$ and M87$^{*}$ offer an important observational probe of strong-field gravity and provide a means of testing theoretical models against astrophysical data. In the following, we analyze the shadows of the PG black hole by adapting the approach developed in Refs.~\citep{lambiase2024weak,pantig2023testing}, we quantify the shadow cast by the black hole in terms of the shadow radius measured by an observer located at radius $r$ as
\begin{eqnarray}
R_{\rm sh}(r_0;r)
=b_0\sqrt{F(r)} 
=r_0 \sqrt{\frac{F(r)}{F(r_0)}}\, ,
\end{eqnarray}
where $b_0=L/E$ is the critical impact parameter associated with the unstable photon orbit at $r=r_0$.
For a distant observer
\begin{equation}
R_{\rm sh}=b_0=\frac{r_0}{\sqrt{F(r_0)}}\,.
\end{equation}

\begin{table*}[t]
\centering
\caption{Fundamental quasinormal mode frequencies ($n=0$, $l=3$) obtained from the eikonal geodesic correspondence and the sixth-order WKB approximation, using the Leaver continued-fraction method as a numerical benchmark. Good agreement is maintained throughout the parameter space, with somewhat larger errors in the torsion-dominated regime where the effective potential becomes flatter. The positive-$q$ case $q=0.8$ exhibits the smallest relative error, $\Delta E=0.002679\%$, consistent with its sharper effective-potential peak (Fig.~\ref{effective_potential}).}
\label{tab:qnm_l3_updated}

\renewcommand{\arraystretch}{1.15}
\setlength{\tabcolsep}{10pt}

\begin{tabular}{c c c c c c c c}
\hline\hline
$q$
& \multicolumn{2}{c}{Null-geodesics}
& \multicolumn{2}{c}{WKB approximation}
& \multicolumn{2}{c}{Leaver method}
& WKB error \\

& ${\rm Re}[\omega]$
& $-{\rm Im}[\omega]$
& ${\rm Re}[\omega]$
& $-{\rm Im}[\omega]$
& ${\rm Re}[\omega]$
& $-{\rm Im}[\omega]$
& $\Delta E\%$
\\
\hline

0.0   & 0.673575  & 0.096225  & 0.675343 & 0.0964872 & 0.675366 & 0.096499& 0.003789 \\
0.20  & 0.697821 & 0.0972209 & 0.699714 & 0.0974722 & 0.699736 & 0.097484 & 0.003533 \\
0.40  & 0.726540 & 0.0979496 & 0.728572 & 0.0981843 & 0.728594 & 0.098195 & 0.003328 \\
0.60  & 0.761783 & 0.0980485 & 0.763956 & 0.0982564 & 0.763977 & 0.098267 & 0.003054 \\
0.80  & 0.807624 & 0.0964726 & 0.809849 & 0.0966322 & 0.809869 & 0.096641 & 0.002679 \\
\hline

-1.5  & 0.560112 & 0.0879708 & 0.561314 & 0.0882578 & 0.561338 & 0.088272 & 0.004908 \\
-3.0  & 0.497110 & 0.0814009 & 0.498038 & 0.0816872 & 0.498063 & 0.081704 & 0.005968 \\
-4.5  & 0.454309 & 0.0762960 & 0.455072 & 0.0765769 & 0.455097 & 0.076594 & 0.006563 \\
-6.0  & 0.422370 & 0.0721865 & 0.423022 & 0.0724608 & 0.423046 & 0.072478 & 0.006879 \\
-7.5  & 0.397173 & 0.0687769 & 0.397744 & 0.0690444 & 0.397768 & 0.069062 & 0.007372 \\
-9.0  & 0.37654 & 0.0658808 & 0.377049 & 0.0661419  & 0.377072 & 0.066159 & 0.007486 \\
-30.0 & 0.249139 & 0.0461385 & 0.249348 & 0.0463413 & 0.249367 & 0.046357 & 0.009717 \\
\hline\hline
\end{tabular}
\end{table*}

\begin{table*}[t]
\centering
\caption{Quasinormal frequencies for higher overtones at $l=15$ in the torsion-dominated regimes ($q=-5$ and $q=-10$). The qualitative effect of torsion remains unchanged as the overtone number increases, suppressing both damping rates and oscillation frequencies. The quoted Pad\'e errors measure the convergence of successive Pad\'e approximants and increase gradually for larger $n$, reflecting the expected reduction in WKB accuracy for higher overtones.}
\label{tab:high_modes_torsion}

\renewcommand{\arraystretch}{1.15}
\setlength{\tabcolsep}{14pt}

\begin{tabular}{c c c c c c c}
\hline\hline

& \multicolumn{3}{c}{$q=-5$}
& \multicolumn{3}{c}{$q=-10$}
\\
\hline
$n$
& ${\rm Re}[\omega]$
& $-{\rm Im}[\omega]$
& Pad\'e error
& ${\rm Re}[\omega]$
& $-{\rm Im}[\omega]$
& Pad\'e error
\\

\hline

0 & 1.96078 & 0.0748484 & 0.000 & 1.61505 & 0.0641871 & 0.000 \\
1 & 1.95736 & 0.2248070 & 0.023 & 1.61190 & 0.1928120 & 0.026 \\
2 & 1.95253 & 0.3764320 & 0.195 & 1.60748 & 0.3230610 & 0.224 \\
3 & 1.94970 & 0.5338180 & 0.774 & 1.60492 & 0.4589720 & 0.891 \\
4 & 1.95263 & 0.7059170 & 2.159 & 1.60744 & 0.6094480 & 2.490 \\
5 & 1.96289 & 0.9101240 & 4.921 & 1.61548 & 0.7920060 & 5.683 \\

\hline\hline
\end{tabular}
\end{table*}

\begin{figure*}[t]
\centering

\begin{minipage}[t]{0.48\textwidth}
\centering
\includegraphics[width=\linewidth]{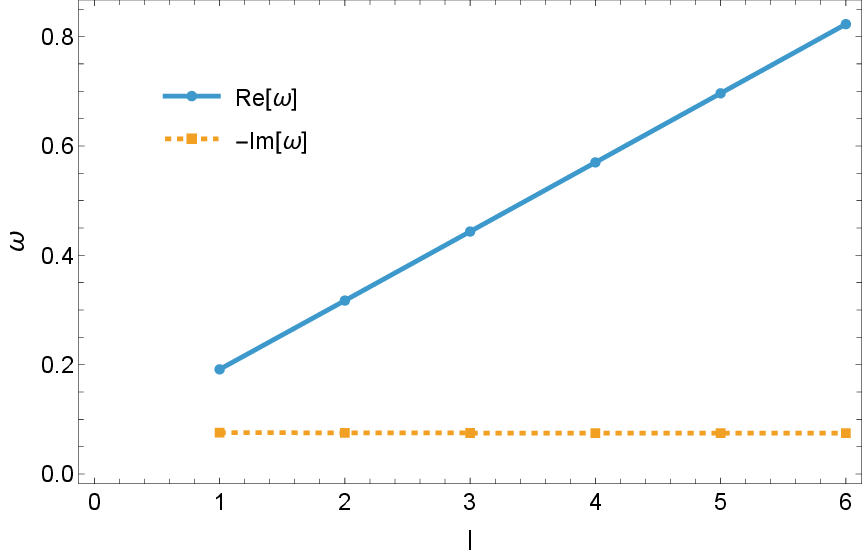}
\\[-2mm]
(a) $q=-5$
\end{minipage}
\hfill
\begin{minipage}[t]{0.48\textwidth}
\centering
\includegraphics[width=\linewidth]{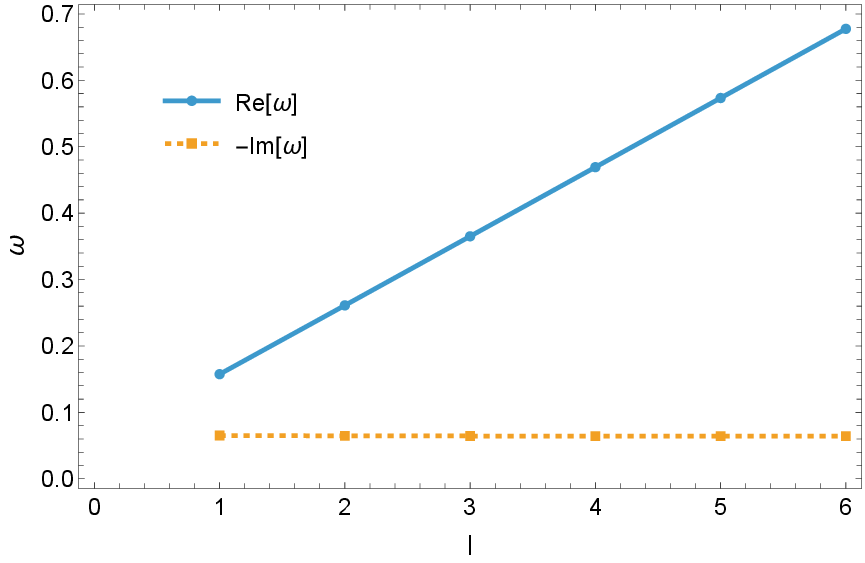}
\\[-2mm]
(b) $q=-10$
\end{minipage}

\caption{Dependence of the fundamental quasinormal mode frequencies on the multipole number $l$ for different values of the torsion parameter. The spectrum exhibits the same qualitative behavior across the parameter space: the oscillation frequency, $\mathrm{Re}[\omega]$, increases monotonically with $l$, while the damping rate, $|\mathrm{Im}[\omega]|$, remains nearly constant. This weak dependence of the imaginary part on the multipole number is consistent with the well-known behavior of black hole quasinormal modes in eikonal regimes, for which variations in $l$ predominantly affect the real part of the frequency.}
\label{fig:omega_vs_l_torsion}
\end{figure*}

The photon-sphere radius follows from the null-geodesic condition in Eq.~\eqref{null}, which admits two solutions, $(r_{0+},r_{0-})$. In the RN-like sector ($q>0$), the inner solution lies inside the event horizon and is therefore physically inaccessible. In the torsion-dominated regime ($q<0$), the inner solution becomes negative and hence unphysical. Consequently, the only physically relevant photon orbit is
\begin{equation}
\label{shad}
r_0=\frac{3m+\sqrt{9m^2-8q}}{2}\,.
\end{equation}
Equation~\eqref{shad} demonstrates that the torsion parameter $q$ directly modifies the location of the photon-sphere and, consequently, the size of the black hole shadow. Black hole shadow observations therefore provide a complementary probe of the torsional sector of the theory, extending the phenomenological insights obtained from the ringdown analysis. For small $q$, expansion about Schwarzschild yields
\begin{equation}
r_0 = 3m - \frac{2}{3}\frac{q}{m}
- \frac{4}{27}\frac{q^2}{m^3} +\mathcal{O}\left(\frac{q^3}{m^5}\right) \, ,
\end{equation}
\begin{equation}
R_{\rm sh} = 3\sqrt{3}m
- \frac{\sqrt{3}}{2}\frac{q}{m}
- \frac{7\sqrt{3}}{72}\frac{q^2}{m^3} +\mathcal{O}\left(\frac{q^3}{m^5}\right)\,.
\end{equation}
Thus,
\begin{equation}
\Delta R \simeq -\frac{\sqrt{3}}{2}\frac{q}{m}\,.
\end{equation}
Furthermore, one can also observe from this relation that the torsion parameter $q$ has a direct effect on the black hole shadow size through the photon sphere radius, where negative values of $q$ increase the size of both the photon sphere and the shadow, while positive values decrease them.

\begin{table*}[t]
\centering
\caption{
Observed angular diameters of M87* and Sgr~A* \citep{Eht1, Eht2}, corresponding Schwarzschild predictions, and constraints on the torsion parameter $q/m^2$. The best-fit values are obtained from a profile-$\chi^2$ analysis including the propagation of uncertainties in the shadow diameter, black hole mass, and source distance. The quoted $1\sigma$ intervals correspond to $\Delta\chi^2=1$ for a single parameter. A Monte Carlo analysis based on the observational uncertainties yields statistically consistent results.
}
\label{tab:shadow_constraints}

\renewcommand{\arraystretch}{1.15}
\setlength{\tabcolsep}{10pt}

\begin{tabular}{l c c c c}
\hline\hline

Source
& $d_{\rm obs}$ ($\mu$as)
& $d_{\rm Schw}$ ($\mu$as)
& $q_{\rm best}/m^2$
& $q_{1\sigma}/m^2$ \\

\hline

M87*
& $42.0 \pm 3.0$
& 39.689
& $-0.373$
& $[-1.705,\,0.421]$ \\

Sgr~A*
& $51.8 \pm 2.3$
& 53.268
& $0.160$
& $[-0.104,\,0.395]$ \\

Combined ($\chi^2$)
& ---
& ---
& $0.114$
& $[-0.140,\,0.341]$ \\

Monte Carlo
& ---
& ---
& $0.037$
& $[-0.237,\,0.312]$ \\

\hline\hline
\end{tabular}
\end{table*}
Following the EHT observations (see Refs. \cite{Eht1,Eht2}), the angular diameter of the black hole shadow is given by
\begin{equation}
\theta(q)=\frac{2R_{\rm sh}(q,M)}{D}\,,
\end{equation}
with
\begin{equation}
M_{\rm M87*}
=(6.5\pm0.2_{\rm stat}\pm0.7_{\rm sys})\times10^{9}M_\odot,
\qquad
D_{\rm M87*}=16.8^{+0.8}_{-0.7}\ {\rm Mpc},
\end{equation}
and
\begin{equation}
M_{\rm SgrA*}
=(4.297\pm0.012\pm0.040)\times10^{6}M_\odot,
\qquad
D_{\rm SgrA*}
=(8277\pm9\pm33)\ {\rm pc}.
\end{equation}
Notably, the EHT angular diameter is identified with the theoretical shadow diameter as a first approximation. Astrophysical effects, including black hole spin, accretion flow structure, and radiative-transfer modeling, may introduce systematic uncertainties; the resulting bounds herein are  therefore indicative.

\begin{figure}[t]
    \centering
    \includegraphics[width=0.75\textwidth]{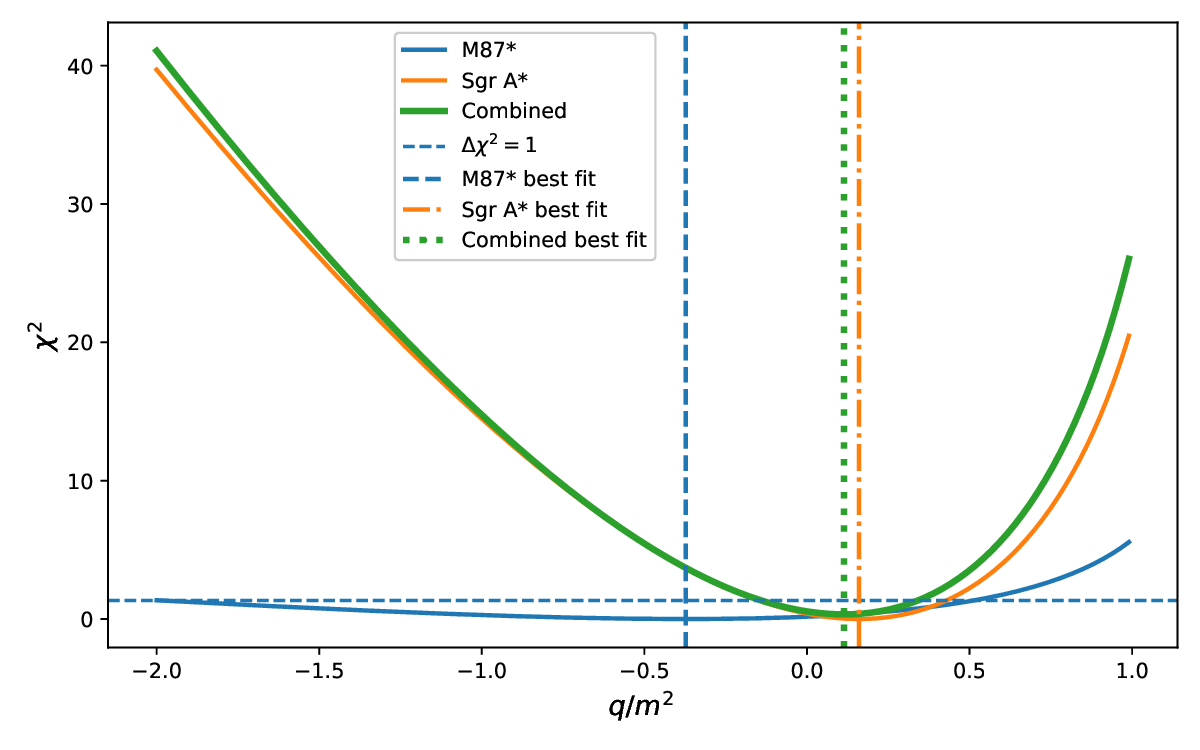}
   \caption{Profiles of the $\chi^2$ statistic as a function of the dimensionless torsion parameter $q/m^2$. The solid blue, orange, and green curves correspond to the likelihood profiles obtained from M87*, Sgr~A*, and their combined analysis, respectively. The horizontal dashed line denotes the $1\sigma$ confidence threshold, corresponding to $\Delta\chi^2=1$ for a single parameter. The vertical blue dashed, orange dot-dashed, and green dotted lines indicate the best-fit values inferred from M87*, Sgr~A*, and the combined dataset, respectively. The individual likelihood profiles are broadly consistent and overlap significantly, yielding the combined constraint
\(
q/m^2 = 0.114^{+0.226}_{-0.254},
\)
at the 68\% confidence level.}
    \label{fchisquare}
\end{figure}

Moreover, the existence of an event horizon requires that
\(
q/m^2 \le 1\,,
\)
and we consider the parameter range
\(
q/m^2\in(-2.0,0.99)\,.
\)
The constraints obtained from the two sources are consistent within uncertainties. As shown in Table \ref{tab:shadow_constraints} and Fig. \ref{fchisquare}, a combined fit, based on a joint $\chi^2$ analysis of the angular-diameter measurements, yields
$
q/m^2 = 0.114^{+0.226}_{-0.254}\,,
$
 at the $1\sigma$ confidence level, where the $1\sigma$ interval is defined by $\Delta\chi^2 < 1$, for a single parameter.

The robustness of this result is further supported by Monte Carlo sampling of the observational uncertainties, which yields consistent best-fit values with a mean $q/m^2 = 0.0371 \in [ -0.237\, , 0.312]$. The agreement between the $\chi^2$  and Monte Carlo estimates indicates that the inferred bounds are stable under statistical fluctuations.

Additionally, the shadow and QNM analyses share a common interpretation: torsion affects various observational features of spacetime by modifying the photon sphere. Since null geodesics govern the behavior of both the shadow radius and the ekinoal QNM spectrum, investigating these observables together serves as a powerful probe of torsion-related deviations from GR. 
\section{Conclusions}
\label{s5}

We have investigated the influence of spacetime torsion on the ringdown spectrum and shadow properties of a black hole in Riemann-Cartan geometry. The solution considered arises in PG theory, where the effective charge is generated by torsion rather than by an electromagnetic field, leading to an RN-like geometry with distinct horizon and observational properties. To study the ringdown dynamics, we employed the sixth-order WKB approximation, the eikonal geodesic correspondence, and Leaver's continued-fraction method as an independent numerical validation.

Our results show that the torsion parameter produces significant modifications of the ringdown spectrum. In particular, the torsion-dominated regime shifts the modes toward lower oscillation frequencies and smaller damping rates, resulting in longer-lived ringdown signals. We traced this behavior to the progressive flattening of the effective potential, which alters both the propagation of perturbations and the properties of unstable null geodesics. To illustrate the possible observational implications of this effect, we further examined the behavior of a spin-2 test field and showed that the corresponding ringdown observables exhibit systematic deviations from their GR counterparts.

A distinctive feature of the torsion-dominated sector is the absence of a physical Cauchy horizon. As the magnitude of the negative effective charge increases, the inner horizon becomes unphysical, and the spacetime no longer admits the horizon degeneracy associated with extremal RN black holes. Since the effective charge originates from torsion rather than from a Maxwell field, the resulting geometry allows a wider range of physically admissible configurations. This property motivates the use of black hole shadows as a complementary probe of the torsional sector. Accordingly, by using the EHT of M87$^{*}$ and Sgr~A$^{*}$, we obtained constraints on the torsion parameter and found that the observationally preferred region remains consistent with the Schwarzschild limit within current uncertainties.

Accordingly, our results demonstrate that torsion modifies the spacetime geometry in a manner that simultaneously affects both ringdown and shadow observables. The same geometric deformation responsible for changes in the quasinormal spectrum also alters the photon-sphere structure and the corresponding shadow size, establishing a direct connection between gravitational-wave and horizon-scale probes of strong gravity. Moreover, despite its RN-like form, the torsion-induced effective charge affects black hole observables differently from a conventional Maxwell charge. Increasing torsion enlarges the photon sphere and shadow while suppressing the quasinormal spectrum, leading to trends opposite to those associated with electrically charged black holes.

Several extensions of the present work merit further investigation. Since the effective charge originates from spacetime torsion rather than from an electromagnetic field, it would be interesting to explore semiclassical processes such as Hawking radiation and greybody factors in this background. A quantum-field-theoretic treatment in Riemann-Cartan spacetime may also reveal additional signatures associated with the coupling between torsion and intrinsic spin. More importantly, a complete analysis of gravitational perturbations obtained by linearizing the full Poincar\'e gauge field equations would clarify the role of dynamical torsion in the ringdown process and provide a firmer basis for future comparisons with observations from LVK, LISA, and next-generation gravitational-wave detectors.
\section*{Data availability}
The WKB calculations reported in this work were performed using the publicly available Mathematica implementation accompanying Ref.~\citep{konoplya2019higher}. The code can be obtained from the authors' repository associated with that work. Additional scripts developed for the numerical calculations presented in this paper are available from the corresponding author upon reasonable request.
\section*{Acknowledgements}
Terkaa Victor Targema and Usman Zafar acknowledge the Japanese government (MEXT) scholarship. The work of Kazuharu Bamba was supported in part by the JSPS KAKENHI
Grants No. 24KF0100 and No. 25KF0176, and a grant-in-aid of academic
research of the Yamaguchi Scholarship Foundation.

\bibliographystyle{apsrev4-2}
\bibliography{example}
\end{document}